\documentclass[runningheads]{llncs}
\usepackage{graphicx}
\usepackage{xcolor}
\usepackage{wrapfig}
\usepackage{algorithm}
\usepackage[noend]{algpseudocode}

\usepackage{bm}
\usepackage{amsmath}
 \usepackage{amssymb}
 \usepackage{caption}
\usepackage{hyperref}
\usepackage[misc,geometry]{ifsym} 
\usepackage{svg}
\newcommand{\orcid}[1]{\href{https://orcid.org/#1}{\includegraphics[width=8pt]{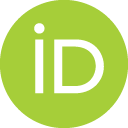}}}

\usepackage{adjustbox}

\begin{document}
\title{Mining Sequential Patterns in Uncertain Databases Using Hierarchical Index Structure
}
\titlerunning{Uncertain seq. mining with hier. index struct.}
%
\author{Kashob Kumar Roy\inst{1}
\orcid{0000-0003-4691-7060}
\and
Md Hasibul Haque Moon\inst{1}
\orcid{0000-0003-4392-6084}
\and
Md Mahmudur Rahman\inst{1}
\orcid{0000-0002-9571-5185} 
\and
Chowdhury Farhan Ahmed \inst{1}\textsuperscript{(\Letter)}
\orcid{0000-0002-6101-4591}
\and
Carson K. Leung\inst{2}
\orcid{0000-0002-7541-9127}
}
\authorrunning{K. K. Roy et al.}
%
\institute{Department of Computer Science \& Engineering, University of Dhaka, Bangladesh
\and
Department of Computer Science, University of Manitoba, Canada\\
\email{kashobroy@gmail.com, hasibulhq.moon@gmail.com, mahmudur@du.ac.bd, farhan@du.ac.bd {\Letter}, kleung@cs.umanitoba.ca}
}

\maketitle              
\begin{abstract}
In this uncertain world, data uncertainty is inherent in many applications
and its importance is growing drastically due to the rapid development of modern technologies. Nowadays, researchers have paid more attention to mine patterns in uncertain databases. A few recent works attempt to mine frequent uncertain sequential patterns. Despite their success, they are incompetent to reduce the number of false-positive pattern generation in their mining process and maintain the patterns efficiently. In this paper, we propose multiple theoretically tightened pruning upper bounds that remarkably reduce the mining space. A novel hierarchical structure is introduced to maintain the patterns in a space-efficient way. Afterward, we develop a versatile framework for mining uncertain sequential patterns that can effectively handle weight constraints as well. Besides, with the advent of incremental uncertain databases, existing works are not scalable. There exist several incremental sequential pattern mining algorithms, but they are limited to mine in precise databases. Therefore, we propose a new technique to adapt our framework to mine patterns when the database is incremental. Finally, we conduct extensive experiments on several real-life datasets and show the efficacy of our framework in different applications.

\keywords{Sequential Pattern Mining \and Uncertain Database  \and Weighted Sequential Patterns \and Incremental Database.}
\end{abstract}
\section{Introduction}
    Sequential Pattern Mining is an important and challenging data mining problem~\cite{pei2004mining_prefixSpan,srikant1996mining_GSP} with broad applications where the order of the itemsets or events in a sequence is important. There are many applications such as environmental surveillance, medical diagnosis, security, and manufacturing systems etc where uncertainty is inherent in nature due to several limitations: (i) our limited understanding
of reality; (ii) limitations of the observation equipment; or (iii) limitations of available resources for the analysis of data, etc. A large number of approaches have been introduced in~\cite{ahmed2016mining_WuncertainFP4,le2020mining,lin2012new_uncertainFP3,lin2016weighted_WuncertainFP5} 
to mine frequent itemsets from uncertain databases. 
Algorithms proposed in~\cite{ge2015mining,zhao2013mining_uncertainSeq} mine sequential patterns in uncertain databases. 
However, in the real world, not all items are equally important.
For example, in biomedical data analysis, some genes are more vital than others in causing a particular disease.
Weighted pattern mining methods are proposed in~\cite{li2020efficient,yun2008new_WSPAN} for this task.
Rahman et al.~\cite{rahman2019mining_uWSeq} 
handle weight constraints in mining uncertain sequential patterns by maintaining weight and expected support threshold separately. Thus, it can efficiently mine sequences having high frequencies with high weights but incompetent to mine sequences which have low frequencies with high weights or high frequencies with low weights. Besides, existing uncertain sequential pattern mining methods have some vital limitations such as: a) generation of a huge number of false-positive patterns due to the pruning upper bounds; b) inefficient maintenance of candidate patterns, which results in costly support computation; and c) lack of a sophisticated weight upper bound to mine weighted patterns efficiently while maintaining anti-monotone property. To address these limitations, we propose multiple novel pruning upper bounds that are theoretically tightened than respective upper bounds already introduced in the literature and utilize a hierarchical index structure to maintain potential candidate patterns in a space-efficient way. 

Moreover, with the advent of modern technologies, most databases are dynamic and incremental in nature. 
A large number of researches~\cite{cheng2004incspan_INCSPAN,ishita2018efficient_WINCSPAN,lyu2019efficient} have been successful in incremental pattern mining.
But none of the existing uncertain sequential pattern mining algorithms are effective in handling the dynamic nature because running batch algorithms from scratch after each increment is not a feasible solution in the sense of time. 
To the best of our knowledge,  our proposed technique is the first work to mine sequential patterns in incremental uncertain databases. In summary, our contributions in this work are as follows,
\begin{enumerate}
    \item Three theoretically tightened upper bounds: $expSup^{cap}$, $wgt^{cap}$, $wExpSup^{cap}$ to reduce the search space of mining potential candidate patterns.
    \item A novel hierarchical index structure, \textit{USeq-Trie}, to maintain the patterns.
    \item A faster method, \textit{SupCalc}, to compute expected support of patterns.
    \item An efficient algorithm, $FUSP$, to mine sequential patterns in uncertain database. 
    \item An approach \textit{InUSP} for incremental mining of uncertain sequential patterns.
\end{enumerate}
Extensive experimental analysis 
validates the efficacy of our proposed methods and shows that our methods consistently outperform other baseline approaches.

\section{Background Study}

\textbf{Related Works.}
Among a plethora of research on sequential pattern mining, \textit{GSP}~\cite{srikant1996mining_GSP} works based on candidate generation and testing paradigm whereas \textit{PrefixSpan}~\cite{pei2004mining_prefixSpan} follows the divide-and-conquer approach to mine frequent sequences in precise databases. \textit{PrefixSpan}~\cite{pei2004mining_prefixSpan} expands patterns by recursively projecting the database into smaller parts and mining local patterns in those prefix-projected databases.
\begin{table}[bt]
\begin{minipage}{0.52\linewidth}
\caption{Initial Database, $DB$}
\centering
\begin{tabular}{|p{0.08\linewidth}|p{0.85\linewidth}|}
\hline
\textbf{Id}& \textbf{Uncertain Sequence} \\ \hline
1 & (a:0.9, c:0.6)(a:0.7)(b:0.3)(d:0.7) \\ \hline
2 & (a:0.6, c:0.4)(a:0.5)(a:0.4, b:0.3) \\ \hline
3 & (a:0.3)(a:0.2, b:0.2)(a:0.4, b:0.3, g:0.5) \\ \hline
4 & (a:0.1, c:0.1)(a:0.3, b:0.1, c:0.4) \\ \hline
5 & (d:0.1)(a:0.4)(d:0.1)(a:0.5, c:0.6) \\ \hline
6 & (b:0.3)(b:0.4)(a:0.1)(a:0.1, b:0.2) \\ \hline
\end{tabular}%
\label{tab:initial_db}

\end{minipage}
\hfill
\begin{minipage}{0.44\linewidth}
\centering
\caption{Weight Table}
\begin{tabular}{|l|l|l|l|}
\hline
\textbf{Item}& \textbf{Weight} & \textbf{Item} &\textbf{Weight}                \\ \hline
a & 0.8 & b & 1.0  \\ \hline
c & 0.9 & d & 0.9 \\ \hline
e & 0.7 & f & 0.9 \\ \hline
g & 0.8 & & \\ \hline

\end{tabular}%
\label{tab:wgt}
\end{minipage}
\end{table}
Uncertain data has gained great attention in recent years~\cite{ahmed2016mining_WuncertainFP4,li2020efficient,muzammal2011mining_uSeq1st,rahman2019mining_uWSeq,zhao2013mining_uncertainSeq}. 
Inspired by \textit{PrefixSpan}, \textit{{U-PrefixSpan}}~\cite{muzammal2011mining_uSeq1st} mines probabilistic frequent sequences whereas
\textit{uWSequence}~\cite{rahman2019mining_uWSeq} mines expected support-based frequent sequences with weight constraints in uncertain databases. 
\textit{uWSequence}~\cite{rahman2019mining_uWSeq} uses $\textit{expSupport\textsuperscript{top}}$ upper-bound to prune the mining space of patterns. They use weight threshold as an extra level of filtering which is not aligned with the concept of weighted support defined in~\cite{yun2008new_WSPAN} for precise databases. Following~\cite{yun2008new_WSPAN}, we introduce the concept of weighted expected support in uncertain sequential pattern mining that considers both expected support and weight of patterns simultaneously.
Further, 
researchers proposed various algorithms in~\cite{cheng2004incspan_INCSPAN,ishita2018efficient_WINCSPAN,lyu2019efficient} to handle increments in databases. \textit{IncSpan}~\cite{cheng2004incspan_INCSPAN} introduces the concept of buffering semi-frequent sequences \textit{(SFS)}
mined from initial databases which may become frequent after future increments. \textit{WIncSpan}~\cite{ishita2018efficient_WINCSPAN} finds weighted sequential patterns in incremental precise databases. Despite the promising significance of incremental uncertain sequential pattern mining in different applications, existing works are not capable to mine patterns efficiently. 
Hence, we introduce a new concept of promising frequent sequences \textit{(PFS)} to improve the efficiency.

\textbf{Preliminaries.}
Let \textit{I = \{ i$_1$, i$_2$,..., i$_n$\}} be the set of all items in a database. 
An event \textit{e$_i$  = (i$_1$, i$_2$,...,i$_k$)} is a subset of \textit{I}. A sequence is an ordered set of events.
For example, \textit{$\alpha$=$<$$(i_2 ),(i_1, i_5),(i_1)$$>$} consists of 3 consecutive events. In uncertain sequences, items in each event are assigned with their existential probabilities such as \textit{$\alpha$ =$<$(i$_2$: P$_{i_2}$), (i$_1$: P$_{i_1}$, i$_5$: P$_{i_5}$), (i$_1$: P$_{i_1}$)$>$}. An uncertain sequential database is a collection of uncertain sequences shown in Table~\ref{tab:initial_db}.
Support of a sequence $\alpha$ in a database is the number of data tuples that contain $\alpha$ as a subsequence. In this paper, we follow the definition of expected support \textit{(expSup)} for a sequence (items within the sequence are independent) which is defined in~\cite{rahman2019mining_uWSeq} as the sum of the maximum possible probabilities of that sequence in each data tuple where the probability of a sequence is computed simply by multiplying the uncertainty value of its all items.
A sequence $\alpha$ can be extended with an item \textit{i} in two ways: 
i) \textit{i-extension}, insert \textit{i} to the last event of $\alpha$, 
and ii) \textit{s-extension}, add \textit{i} to $\alpha$ as a new event. 
{Weight of a sequence} \textit{(sWeight)} is the sum of its each individual item’s weight divided by the length of the sequence \cite{yun2008new_WSPAN} i.e., the total number of items in the sequence. 
According to Table~\ref{tab:initial_db} and Table~\ref{tab:wgt}, for sequence $\alpha$ = $<$(a)(b)$>$, support of $\alpha$ is 5, $expSup(\alpha)= max(0.9\times0.3, 0.7\times0.3) 
+ max(0.6\times0.3, 0.5\times0.3) 
+ max(0.3\times0.2, 0.3\times0.3, 0.2\times0.3) 
+ (0.1\times0.1)
+ 0 
+ (0.1\times0.2) = 0.57$,
and $sWeight(\alpha) = (0.8+1.0)/2 = 0.9$ as per the definitions.

\section{A Framework for mining Uncertain Sequential Patterns}
In this section, we propose a new framework for mining sequential patterns in uncertain databases efficiently with/without the weight constraints in mining patterns followed by discussing the incremental mining approach when the database would be of dynamic nature.

\textbf{Definitions.} $maxPr$ is the maximum possible probability of a sequence $\alpha = <(i_{1})(i_{2})...(i_{|\alpha|})>$ in the 
whole
database \cite{rahman2019mining_uWSeq},
\begin{equation}
    {maxPr(\alpha)} = \prod^{|\alpha|}_{k = 1} (\widehat{P}_{DB\mid\alpha_{k-1}}(i_{k}))\ where\ \alpha_{k-1} = <(i_{1})...(i_{k-1})>
\end{equation}
where $\widehat{P}_{DB\mid \alpha}(i)$ = maximum possible probability of item $i$ in a database  \textit{$DB\mid \alpha$} that is the projection of original database with $\alpha$ as current prefix~\cite{pei2004mining_prefixSpan}. Moreover, \cite{rahman2019mining_uWSeq} shows that the \textit{maxPr} measure holds anti-monotone property. Similar to \textit{maxPr}, we define another measure $maxPr_{S}(\alpha)$ as the maximum probability of a pattern $\alpha$ in a single data sequence \textit{S}.
According to Table \ref{tab:initial_db}, the \textit{maxPr}($<(c)(a)>$) = $0.6 \times 0.7$ = 0.54 and \textit{maxPr}($<(ac)>$) = $0.9 \times 0.6$ = 0.54; where for the 1st data sequence, $maxPr_{S}$($<$(a)(b)$>$) = $max$($0.9 \times 0.3$, $0.7 \times 0.3$) = 0.27. \\ 
We define an upper bound of expected support  of a sequence $\alpha$ of length m as, 
 \begin{equation}
     expSup^{cap}(\alpha_{m})= maxPr(\alpha_{m-1})\times\sum_{\forall S\in (DB|{\alpha_{m-1}})} maxPr_{S}(i_{m})
 \end{equation}


\begin{lemma}
For a sequence $\alpha$, $expSup^{cap}(\alpha)\geq expSup(\alpha)$ and $expSup(\alpha)\geq expSup(\alpha^{'})$,where $\alpha \subseteq \alpha^{'}$; $\therefore$ $expSup^{cap}(\alpha)\geq expSup(\alpha^{'})$. If $expSup^{cap}(\alpha)$ $<$ a minimum threshold $\gamma$ holds, then $expSup(\alpha)<\gamma$ and $expSup(\alpha')<\gamma, \forall \alpha^{'} \supseteq \alpha$ must be true. 
Thus it satisfies the anti-monotonicity constraints.
\label{lem:es01}
\end{lemma}
\begin{lemma}
For a sequence $\alpha$, $expSup^{cap}(\alpha)$ $\leq$ $expSupport^{top}(\alpha)$\footnote{uWSequence\cite{rahman2019mining_uWSeq} defines the upper bound of expected support as $expSupport^{top}(\alpha)$ = $maxPr(\alpha_{m-1})\times maxPr(i_{m}) \times sup_{i_{m}}$ where $sup_{i_{m}}$ is the support count of $i_{m}$.}always holds. Hence, $expSup^{cap}(\alpha)$ significantly reduces the search space in mining patterns and leads to a smaller number of false positive patterns than $expSupport^{top}(\alpha)$. 
\label{lem:03_capLessTop}
\end{lemma}
Later on, we define few more definitions where each item has a weight to indicate its importance. We will be consistent with weighted pattern mining in following sections. Note that our framework is easily adaptable to mine patterns without weight constraints that is discussed in the experiments section. 
Following the concept of \textit{weighted support} for precise database in~\cite{yun2008new_WSPAN}, 
we define \textit{weighted expected support} of a sequence $\alpha$ as
$WES(\alpha) = expSup(\alpha) \times sWeight(\alpha)$. According to Tables~\ref{tab:initial_db} and \ref{tab:wgt}, $WES$($<$(a)(b)$>$) = $0.57\times0.9$ = $0.513$.  
A sequence $\alpha$ is called \textit{weighted sequential pattern} if $WES(\alpha)$ meets a minimum threshold. This threshold is defined to be
$minWES = min\_sup\times (size\ of\ the\ whole\ database)\times WAM \times wgtFct$. 
Here, $min\_sup$ is user given value in range [0,1] related to a sequence's frequency, $WAM$ is weighted arithmetic mean of all item-weights present in the database and defined as
$WAM = {(\sum_{i\in I}^{} w_{i}\times f_{i})}/{\sum_{i\in I}^{} f_{i}}$, 
where \textit{w$_{i}$} and \textit{f$_{i}$} are the weight and frequency of item \textit{i} in current database. Hence, the value of \textit{WAM} changes after each increment in the database. $wgtFct$ is a user-given positive value chosen to tune the mining of weighted sequential patterns. Choice of $min\_sup$ and $wgtFct$ depends on how much frequent and weighted patterns are required in the respective applications.

However, the measure $WES$ does not hold anti-monotone property as any item with higher weight can be appended to a weighted-infrequent sequence and the resulting super-sequence may become weighted-frequent. So, to employ anti-monotone property in mining weighted frequent patterns, we propose two other upper bound measures, $wgt^{cap}$  and $wExpSup^{cap}$, which are used as upper bound of \textit{weight} and \textit{weighted expected support} respectively. Upper bound of weight of a sequence $\alpha$, $wgt^{cap}(\alpha)$ is defined as, 
 \begin{equation}
    wgt^{cap}(\alpha) = \max (mxW_{DB}(DB|\alpha),  mxW_{s}(\alpha))
    \label{eq-wghtcap}
\end{equation}
where $mxW_{DB}(DB|\alpha)$ is the \textit{maximum weight of all frequent items in the $\alpha$-projected database} and $mxW_{s}(\alpha)$ is the \textit{maximum weight of all items in the sequence $\alpha$}. 
To enforce the anti-monotone property of weighted frequent patterns in precise databases,
authors in~\cite{ishita2018efficient_WINCSPAN,yun2008new_WSPAN} make an attempt to use the maximal weight of all items in database as upper bound of weight of a sequence. 
It is obvious to see that $wgt^{cap}$ of a sequence is always less than or equal to the maximal weight of all items in database.
As $wgt^{cap}$ becomes tighter, it generates fewer false positive patterns compared to the existing methods. 

\begin{lemma} 
For any sequence $\alpha$, $wgt^{cap}(\alpha)$ is at least equal to the $sWeight$ value of $\alpha$ and all of its supersequences, $\alpha'$.
Because, $wgt^{cap}(\alpha) \geq sWeight(\alpha)$ and $wgt^{cap}(\alpha) \geq wgt^{cap}(\alpha^{'})$, where $\alpha \subseteq \alpha^{'}$; $\therefore wgt^{cap}(\alpha) \geq sWeight(\alpha^{'})$.
\label{lem:mw}
\end{lemma}
%
%
The proposed upper bound of weighted expected support
is defined as,
\begin{equation}
    {wExpSup^{cap}(\alpha)} = expSup^{cap}(\alpha) \times wgt^{cap}({\alpha})
    \label{eq_wExpSup_cap}
\end{equation}
\begin{lemma}
\label{lem:wescap}
For a sequence $\alpha$, if $wExpSup^{cap}(\alpha)< minWES$, then none of $\alpha$ and its supersequences can be weighted frequent.
Because, $wExpSup^{cap}(\alpha) \geq WES(\alpha)$, and $wExpSup^{cap}(\alpha) \geq WES(\alpha^{'})$, for all $\alpha \subseteq \alpha^{'}$.
\label{lemma_pruning}
\end{lemma}
According to Lemma \ref{lemma_pruning}, we can safely define our pruning condition to reduce the search space of patterns in pattern-growth based mining as follows:

\textit{If for any k-sequence $\alpha$, $wExpSup^{cap}(\alpha)<minWES$, then searching possible extension of $\alpha$ to (k+1)-sequence can be pruned, i.e, neither $\alpha$ nor any super sequences of $\alpha$ would be frequent at all.}

Moreover, Lemma~\ref{lemma_pruning} ensures that our proposed algorithms do not generate any false negative patterns. However, as $wExpSup^{cap}(\alpha)\geq WES(\alpha)$, some patterns may be discovered with $wExpSup^{cap}(\alpha)\geq minWES$ but $WES(\alpha)<minWES$. 
An extra scan of the database is required to remove them.
We have omitted proof of the lemmas due to space limitation.

\subsection{USeq-Trie: Maintenance of Patterns}
We use a hierarchical data structure, named as \textit{USeq-Trie}, to store uncertain sequences and  update their weighted expected support efficiently.
%
%
%
\begin{figure}[b]
\centering
\includegraphics[width=.9\linewidth, height=.41\linewidth]{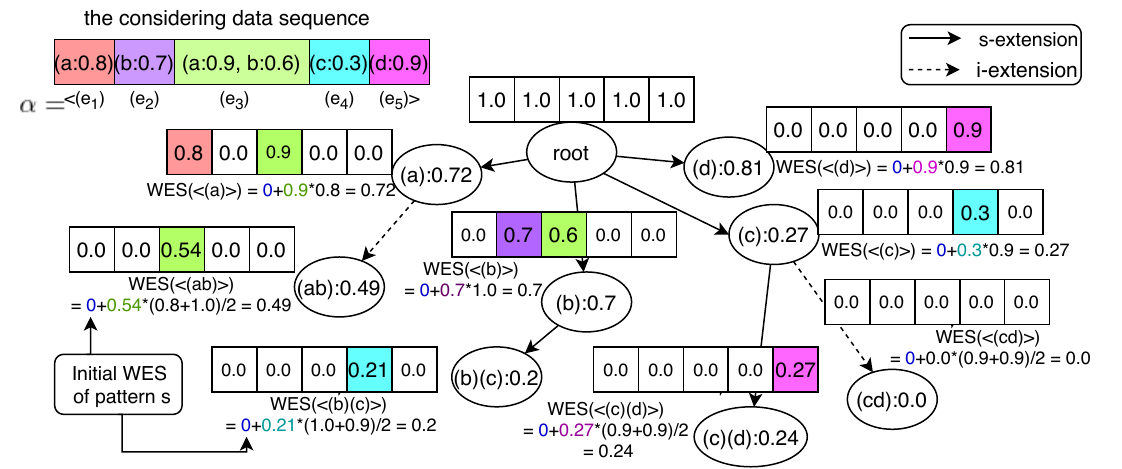}
\caption{An efficient way to compute \textit{WES} of patterns stored into \textit{USeq-Trie}}
\label{fig:useq-trie}
\end{figure}
Each node in the \textit{USeq-Trie} represents an item in a sequence and will be created as either \textit{s-extension} or \textit{i-extension} from its parent node.
Recall that a sequence is an ordered set of events, and an event is a set of items.
In \textit{s-extension}, the edge label is added as a different event. 
In \textit{i-extension}, it is added in the same event as its parent. Each edge is labeled by an item. 
The edge labels in a path to from root to a node forms a pattern. For example, $<$(a)$>$, $<$(b)$>$, $<$(ab)$>$, $<$(c)$>$, $<$(b)(c)$>$, $<$(d)$>$, $<$(cd)$>$ and $<$(c)(d)$>$ are sequential patterns who are stored into \textit{USeq-Trie} shown in Fig.~\ref{fig:useq-trie}. In this figure,  the \textit{s-extensions} are denoted by the \textit{solid lines} and \textit{i-extensions} by \textit{dashed lines}. 
For simplicity of the figure, we are not showing edge labels here. 
Each node represents a (weighted) frequent uncertain sequence and stores its (weighted) expected support.
Now, we present an efficient method, \textit{SupCalc}, to calculate $expSup$ or $WES$ for each candidate pattern stored in a \textit{USeq-Trie}. 

\textbf{Support Calculation, SupCalc}. It reads sequences from the dataset one by one and updates the support of all patterns in \textit{USeq-Trie} against them. For a sequence $\alpha = <e_{1}e_{2}..e_{n}>$ (where $e_{i}$ is an event/itemset), the steps are following,
\begin{enumerate}
    \item Define an array of size \textit{n} at each node. For the root node, all values are 1.0.
    At a particular node, the maximum expected support of pattern \textit{s} from root to that node is stored at proper indices of the node's array - are the ending positions of \textit{s} as a sub-sequence in $\alpha$. The values at other indices are 0.0. 
    \item While traversing the \textit{USeq-Trie} in depth-first order: (i)
    For a node created by a
    \textit{s-extension} with an item \textit{$i_{k}$}, we iterate over all events in $\alpha$ and calculate the support of the current pattern \textit{s} (ends with $i_{k}$ in a new event) by multiplying the probability of item \textit{$i_{k}$} in current event $e_{m}$ with the maximum probability in the parent node's array up to the event $e_{m-1}$. The resulting support is stored at position \textit{m} in the following node's array. 
    (ii) For \textit{i-extension}, the support will be calculated by multiplying the probability of the item \textit{$i_{k}$} in $e_{m}$ with the value at position m in the parent node's array and stored at position \textit{m} in the following child node's array. 
    After that, the maximum value in the resulting array multiplied by its weight will be added to the weighted expected support of the current pattern at the corresponding node.
    \item Use the resultant array to calculate the weighted expected support of all super patterns while traversing the next child nodes.
\end{enumerate}
Fig.~\ref{fig:useq-trie} shows the resulting \textit{USeq-Trie} after updating \textit{WES} for all the stored patterns against a sequence, \textit{$\alpha$=$<$(a:0.8)(b:0.7)(a:0.9,b:0.6)(c:0.3)(d:0.9)$>$}.

\textbf{Complexity of \textit{SupCalc}.}
It takes $O(N\times |\alpha|)$ for updating \textit{N} number of nodes against the sequence $\alpha$. Therefore,  the total time complexity of actual support calculation is $O(|DB|\times N\times k)$ where \textit{k} is the maximum sequence length in the dataset. It outperforms the procedure used in \textit{uWSequence}~\cite{rahman2019mining_uWSeq}
which needs $O(|DB|\times N\times k^{2})$ to calculate a sequence's actual expected support.
Moreover,  we can remove false-positive patterns and find frequent ones from the \textit{USeq-Trie} in \textit{O(N)}. 
Thus, the use of \textit{USeq-Trie} has made our method efficient.
\subsection{FUSP: Faster mining of Uncertain Sequential Patterns}
\label{fuws_proposed}
Inspired by \textit{PrefixSpan}~\cite{pei2004mining_prefixSpan}, we propose \textit{FUSP} to mine weighted sequential patterns in an uncertain database.
It uses the $wExpSup^{cap}$ measure and $SupCalc$ method to reduce the search space and improve the efficiency. 
The sketch of \textit{FUSP} algorithm  is as follows.

\begin{enumerate}
    \item Process the database such that the existential probability of an item in a sequence is replaced with the maximum probability of all of its next occurrences in this sequence. This idea is similar to the \textit{preprocess} function of \textit{uWSequence} \cite{rahman2019mining_uWSeq}. This preprocessed database will be used to run the \textit{PrefixSpan}-like mining approach to find the candidates for frequent sequences.
    While processing, sort the items in an event/itemset in lexicographical order.
    \item Calculate \textit{WAM} of all items present in the current database and calculate the threshold of weighted expected support,  \textit{minWES}.
    \item Find length-1 frequent items and for each item,  project the preprocessed database into smaller parts and expand longer patterns recursively. Store the candidates into a \textit{USeq-Trie}.
    \item While growing longer patterns,  extend current prefix $\alpha$ to $\alpha'$ with an item $\beta$ as  \textit{s-extension} or \textit{i-extension} according to the pruning condition.
    \item Use of $wExpSup^{cap}$ value instead of actual support generates few false-positive candidates. Scan the whole actual database,  update weighted expected supports and prune false-positive candidates based on their \textit{WES}.
\end{enumerate}

\subsection{InUSP: Incremental mining of Uncertain Sequential Patterns}
Existing incremental works~\cite{cheng2004incspan_INCSPAN,ishita2018efficient_WINCSPAN} follow the technique to lower the minimum support threshold by a user-given buffer ratio, $\mu\in [0,1]$, and find \textit{almost frequent} sequence called \textit{SFS} - stating that most of the frequent patterns in the appended database will either come from \textit{SFS} or already frequent sequences (\textit{FS}) in the initial database. Inspired by this concept, we use $minWES^{'}=minWES\times \mu$ to find \textit{SFS} where $minWES^{'}$ $\leq$ \textit{WES} $<$ \textit{minWES}, along with \textit{FS} where \textit{WES} $\geq$ \textit{minWES}. 
However, we argue that \textit{SFS} is not necessarily enough to capture new frequent patterns in future increments.
Let us consider some cases: (a)  an increment to the database may introduce a new sequence which was initially absent in both \textit{FS} and \textit{SFS} but frequently appeared in later increments; (b) a sequence had become infrequent after an increment but could have become semi-frequent or even frequent again after next few increments. There are many real-life cases where new frequent patterns might appear in future increments due to its seasonal behavior or different other characteristics. Existing approaches do not handle these cases. 
To address these cases, we propose to maintain another set of sequences denoted as \textit{Promising Frequent Sequences} (\textit{PFS}) which are neither globally frequent nor semi-frequent after each increment $\Delta DB$  introduced into \textit{DB} but their \textit{WES} satisfy a user-specified threshold that can be defined as $LWES = \gamma \times \mu \times min\_sup\times |\Delta DB| \times WAM \times wgtFct$ where $\gamma$ is a constant factor, to find locally frequent patterns in $\Delta DB$  at a particular point. Here, the globally frequent or semi-frequent implies when considering the size of the
entire database, and locally frequent when using the size of only one increment. Intuitively, we can say that locally frequent patterns may become globally frequent or semi-frequent after next few increments. The patterns whose \textit{WES} values do not meet the local threshold \textit{LWES}, are very unlikely to become globally frequent or semi-frequents. Thus maintaining \textit{PFS} may significantly increase the performance of an algorithm in finding the almost complete set of frequent patterns after each increment. Therefore, we devise \textit{InUSP} to incorporate the concept of \textit{PFS} in mining patterns. Instead of performing \textit{FUSP} from scratch after each increment, \textit{InUSP} works only on $\Delta DB$. Initially, it runs \textit{FUSP} once to find out \textit{FS} and \textit{SFS} from initial database and uses \textit{USeq-Trie} to store \textit{FS} and \textit{SFS}. In addition, a different \textit{USeq-Trie}, which is initially empty, is used to store \textit{PFS}.

After each increment $\Delta DB$,  the steps of $InUSP$ algorithm are as follows:
\begin{enumerate}
    \item Update the values of \textit{database size},  \textit{WAM}, \textit{minWES}, and \textit{$minWES^{'}$}.
    \item Run \textit{FUWS} only in $\Delta DB$ to find locally frequent sequences (\textit{LFS}) against a local threshold,  \textit{LWES}, and store them into \textit{USeq-Trie}. Users can choose \textit{LWES} based on the aspects of application. 
    \item For all $\alpha$ in \textit{FS},  \textit{SFS} and \textit{PFS},  update \textit{WES$_{\alpha}$} using the \textit{SupCalc} method.
    \begin{itemize}
        \item if $WES_{\alpha} <LWES$,  delete $\alpha$'s information.
        \item else if $WES_{\alpha} <minWES'$,  move $\alpha$ to $PFS'$. 
        \item else if $WES_{\alpha} < minWES$, move $\alpha$ to $SFS'$.
        \item else move $\alpha$ to $FS'$.
    \end{itemize}
    \item Move new patterns $\alpha$ from \textit{LFS} to $PFS'$ or  $SFS'$ or $FS'$ based on \textit{WES$_{\alpha}$}.
    \item Use $FS'$,  $SFS'$,  and $PFS'$ as \textit{FS},  \textit{SFS},  and \textit{PFS} respectively for the next increment.
\end{enumerate}

\section{Experimental Results}
We have evaluated our algorithms using several real-life and popular datasets such as \textit{Sign, Kosarak, Fifa, Leviathan, Retail, Foodmart, Chainstore}, and \textit{Online Retail}
from \textit{SPMF}\footnote{\url{http://www.philippe-fournier-viger.com/spmf/index.php?link=datasets.php}} data repository.
We assigned probability and weight values to the items of these datasets as all of them were precise and none of them contained weight information.
We followed normal distribution with \textit{mean} of \textit{0.5} and \textit{standard deviation} of \textit{0.25 (for probabilities)} or \textit{0.125 (for weights)} to generate these values. 
We implemented our algorithms in \textit{Python} programming language and a machine with \textit{Core™ i5-9600U 2.90GHz CPU} and \textit{8GB RAM}.\\

\begin{figure}[tb]
\centering
\includegraphics[width=.6\linewidth, height=.3\linewidth]{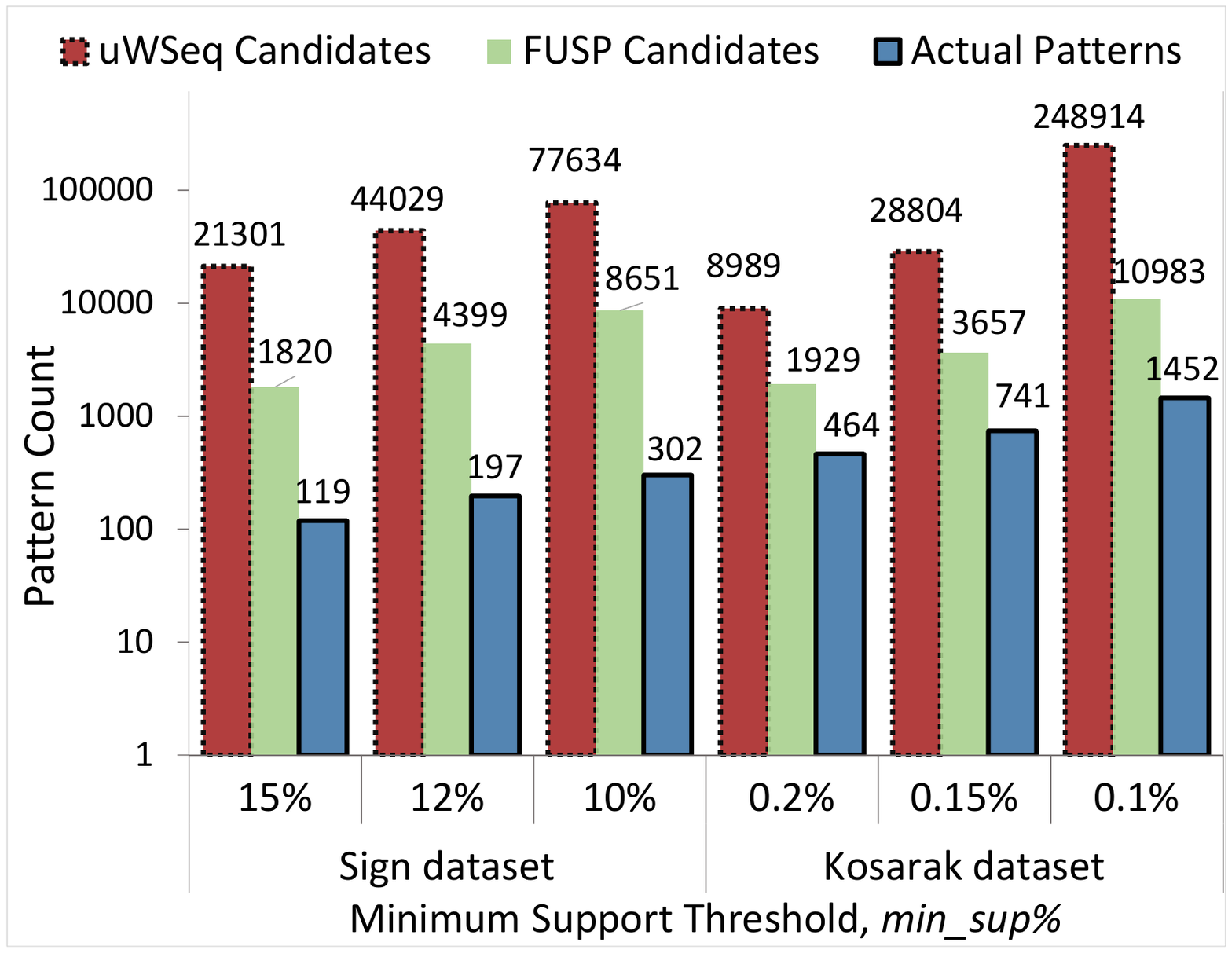}
			\caption{\textit{FUSP} outperforms \textit{uWSequence} in candidate generation}
			\label{fig:cand-fusp}
\end{figure}
\begin{table}[tb]
\centering
\caption{Runtime (seconds) comparison between \textit{uWSequence} and \textit{FUSP}}
\label{tab:fusp-rt}
\begin{tabular}{|ccr|ccr|ccr|}
\hline
\multicolumn{3}{|c|}{\textbf{Sign Dataset}} & \multicolumn{3}{c|}{\textbf{Kosarak Dataset}} & \multicolumn{3}{c|}{\textbf{Fifa Dataset}} \\ \hline
\textit{min\_sup} & uWSeq. & \multicolumn{1}{c|}{FUSP} & \textit{min\_sup} & uWSeq. & \multicolumn{1}{c|}{FUSP} & \textit{min\_sup} & uWSeq. & \multicolumn{1}{c|}{FUSP} \\
\textit{20\%} & 717.69 & 10.64 & \textit{0.25\%} & 5942.06 & 348.32 & \textit{20\%} & 1615.50 & 12.73 \\
\textit{18\%} & 1116.75 & 18.34 & \textit{0.22\%} & 7102.27 & 443.13 & \textit{18\%} & 2943.45 & 25.85 \\
\textit{15\%} & 2052.04 & 32.64 & \textit{0.2\%} & 8581.56 & 475.12 & \textit{17\%} & 4003.97 & 34.79 \\
\textit{12\%} & 4316.43 & 72.39 & \textit{0.18\%} & 14622.38 & 659.30 & \textit{16\%} & 6114.34 & 56.05 \\
\textit{10\%} & 7275.41 & 122.94 & \textit{0.15\%} & 33864.18 & 1029.70 & \textit{15\%} & 9033.86 & 74.95 \\ \hline
\end{tabular}
\end{table}
\textbf{Performance of \textit{FUSP}.}
We have compared with the recent algorithm, \textit{uWSequence}~\cite{rahman2019mining_uWSeq}, which  proposed a framework where the definition of weighted sequential pattern in uncertain databases is different from ours. 
Furthermore, \textit{uWSequence}~\cite{rahman2019mining_uWSeq} outperforms existing methods for mining sequential patterns also without weight constraints in uncertain databases.
So, to show the efficiency of \textit{FUSP} in mining uncertain sequential patterns without weight constraints, we have compared \textit{FUSP} with the current best \textit{uWSequence} by setting the weights of all items to 1.0 which brings both algorithms under a unifying framework.\\
\textbf{(a) False Candidate Generation: }
Recall that both \textit{FUSP} and
\textit{uWSequence} work like \textit{PrefixSpan} using some upper bound of actual expected support value and thus, generate some false positive candidates.
From Fig.  \ref{fig:cand-fusp}, we can see that \textit{FUSP} generates a smaller number of false candidates for any support threshold as it uses a tighter upper bound.
For example, in the \textit{Sign} (dense) dataset with 15\% minimum support threshold, it generates 11 times fewer candidates compared to \textit{uWSequence}.
In \textit{Kosarak} (sparse) with 0.15\% support threshold, \textit{FUSP} generates only 79.7\% false candidates where for \textit{uWSequence}, it is 97.4\%.\\
\textbf{(b) Runtime Analysis:}
\textit{FUSP} needs to maintain a smaller number of candidate patterns in its mining process and uses a faster method to calculate expected support of a pattern. 
Thus, it is a way faster than the \textit{uWSequence} for any support threshold.
Results shown in Table \ref{tab:fusp-rt} validates this claim.
We can see \textit{FUSP} is 50-70 times faster in \textit{Sign} dataset for different thresholds.
Interestingly, the difference in their runtime increases with the decrease in the threshold parameter.
We have found similar results also in other datasets.\\

{\textbf{Performance of the Incremental Technique, \textit{InUSP}.}}
We have modified the current best incremental solution, \textit{WIncSpan} \cite{ishita2018efficient_WINCSPAN} to work in uncertain data by replacing the core PrefixSpan-like algorithm by \textit{FUSP} so that both the proposed \textit{InUSP} and modified $WIncSpan'$ mine weighted sequential patterns from uncertain database.
The baseline approach is running \textit{FUSP} from scratch in the whole updated database after each increment.
We define completeness of the result from an incremental solution to be the percentage of patterns found with respect to the result of the baseline.
To use the datasets as incremental ones, we used the first 50\% of the dataset to be the initial part and then introduced 5 increments of random sizes\footnote{For the \textit{Retail} market-basket dataset, we used the first one-fifth transactions (1st month) as the initial portion and then 4 increments to represent the next 4 months.}, unless mentioned otherwise. 
\\
\begin{figure}[bt]
\centering
		\begin{minipage}[t]{0.47\linewidth}
		\centering
		\includegraphics[width=.9\textwidth, height=.6\textwidth]{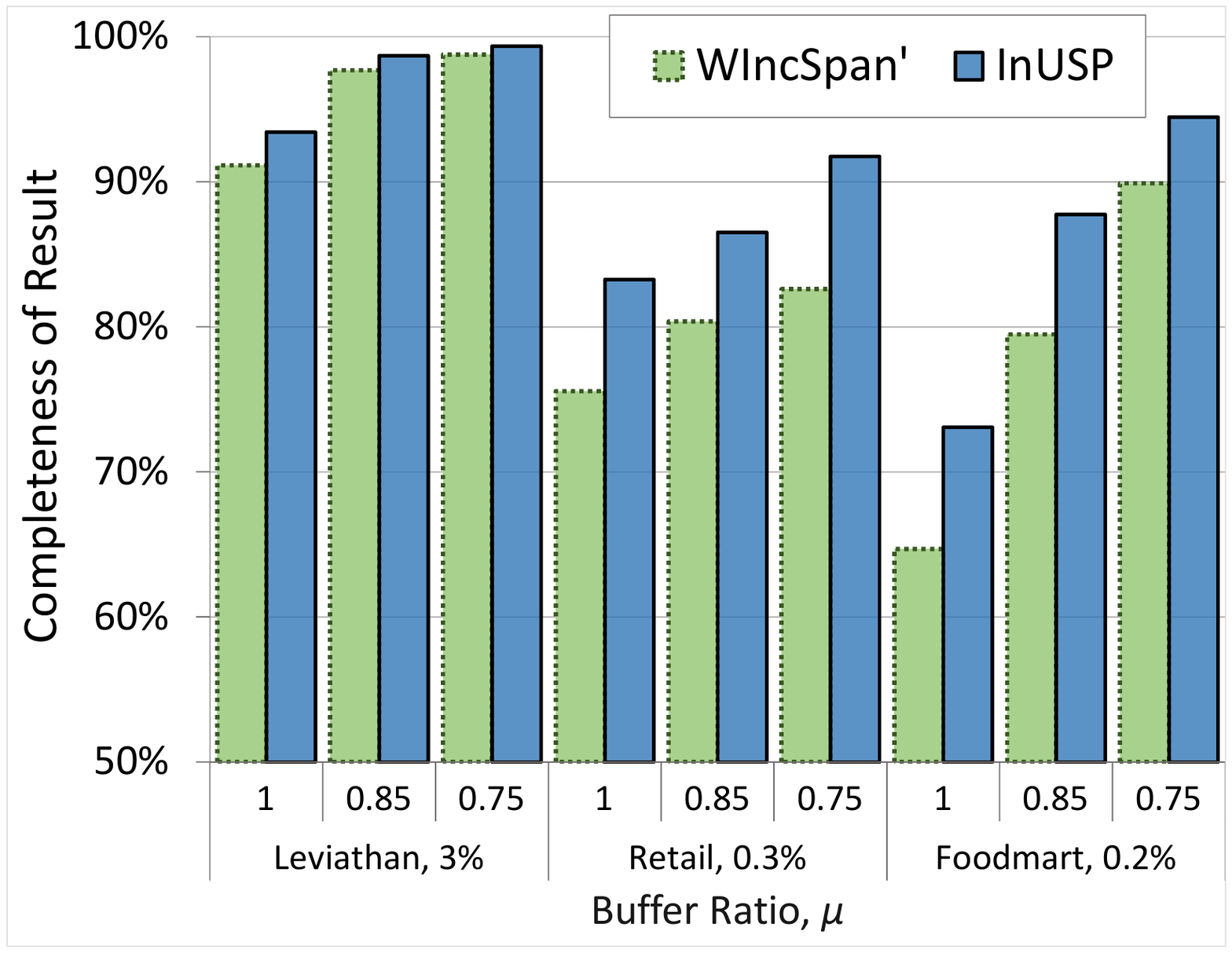}
			\caption{Completeness comparison between $WIncSpan'$ and \textit{InUSP} 
			}
			\label{pat_mu}
	\end{minipage}
	\hfil
	\begin{minipage}[t]{0.47\linewidth}
    \centering
		\includegraphics[width=.9\textwidth, height=.6\textwidth]{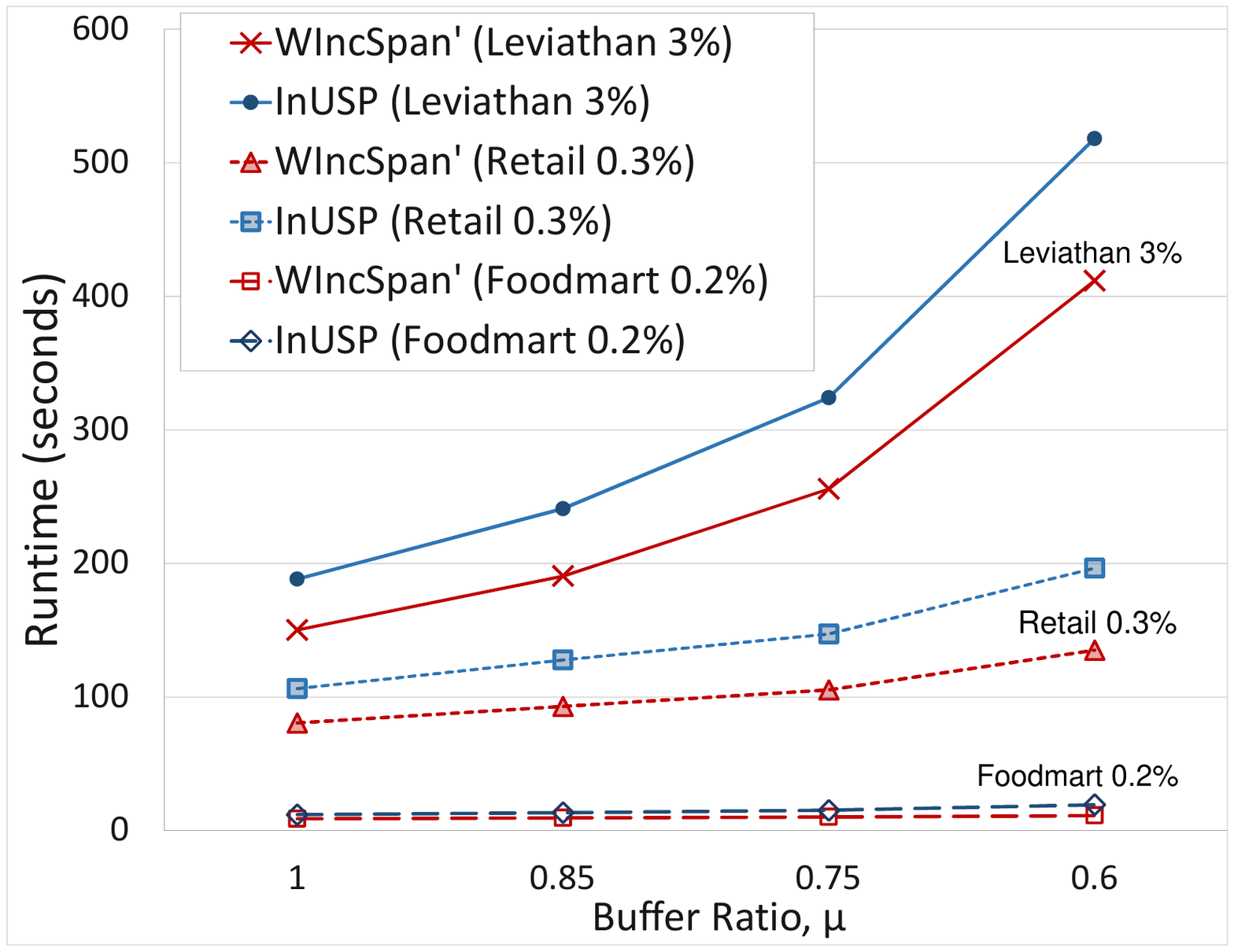}
			\caption{Runtime comparison between $WIncSpan'$ and proposed \textit{InUSP} 
			}
			\label{rt_mu}
	\end{minipage}
\end{figure}
\textbf{(a) Analysis with respect to buffer ratio: }
Buffer ratio,
$\mu=1.0$ means no buffer and lower values mean larger buffers to store semi-frequent sequences. 
Thus, with lower $\mu$, incremental approaches generate and maintain more patterns which help to increase the completeness of their result.
However, due to local mining in incremented portions and maintaining additional promising sequences, \textit{InUSP} always achieves more completeness than $WIncSpan'$.
For the same reason, it also requires slightly more time than $WIncSpan'$.
From Fig. \ref{pat_mu} and Fig. \ref{rt_mu}, we can see the trade-off between completeness and runtime.
We observe that difference in completeness is larger in datasets like \textit{Retail} and \textit{Foodmart} (market-basket) where increments contain frequent items or introduce new items frequently than datasets like \textit{Leviathan} (word sequences) where the initial database contains almost all of the frequent sequences.
By repeating this experiment in other datasets and by varying the support threshold, 
we find that though \textit{InUSP} consumes slightly more time, it outperforms $WIncSpan'$ in terms of completeness of result in every dataset for any combination of $\mu$ and $min\_sup$.\\
\begin{figure}[tb]
\centering
    	\begin{minipage}[t]{0.47\linewidth}
    	\centering
		\includegraphics[width=.9\textwidth, height=.6\textwidth]{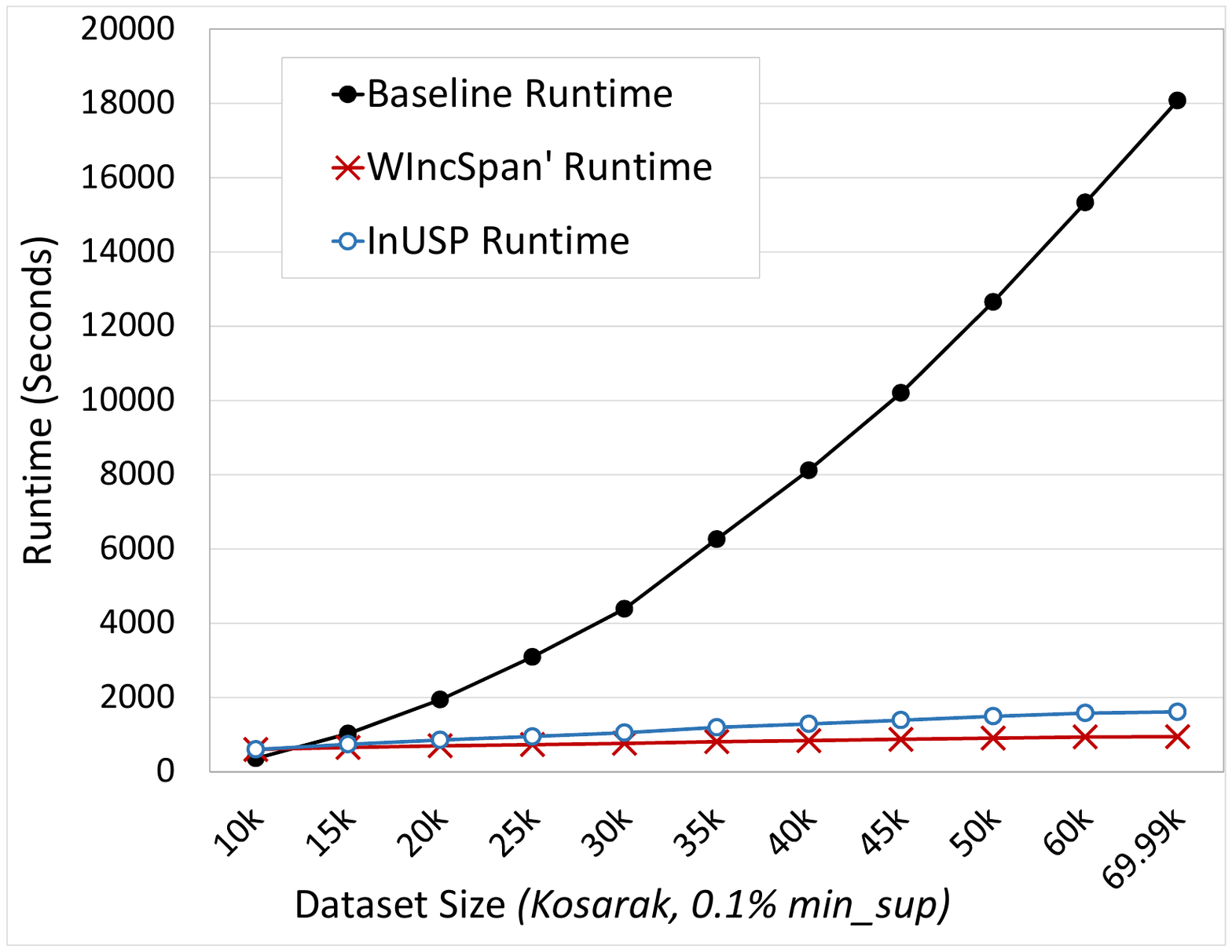}
			\caption{Comparison of scalability using \textit{Kosarak} dataset}
			\label{scale}
	\end{minipage}
    \hfil
		\begin{minipage}[t]{0.47\linewidth}
        \centering
		\includegraphics[width=.9\textwidth, height=.6\textwidth]{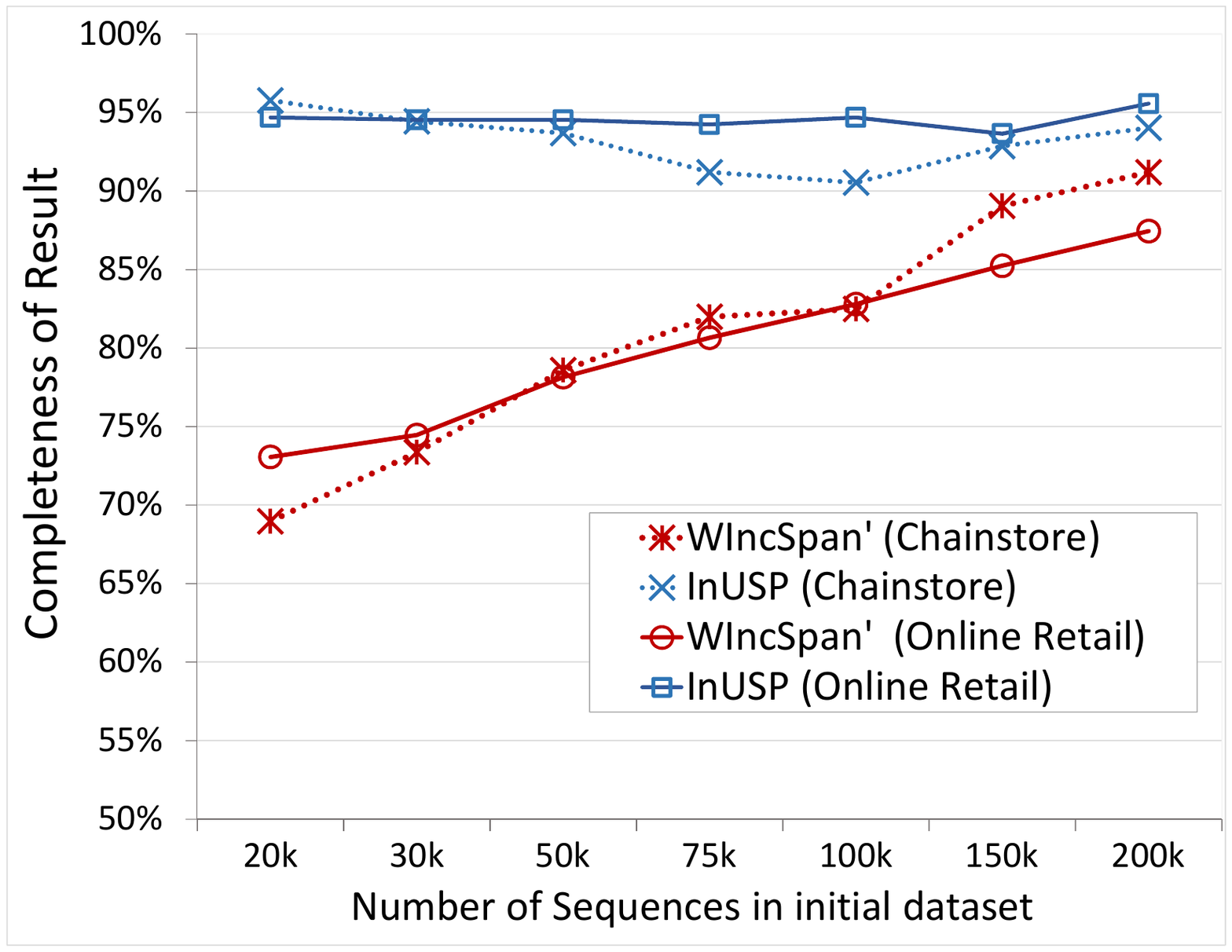}
			\caption{Change in completeness for different initial sizes of a dataset 
			}
			\label{comp_init}
	\end{minipage}

\end{figure}
\textbf{(b) Scalability Analysis: }
To test scalability we have run \textit{InUSP, $WIncSpan'$} and the baseline approach in several large datasets introducing several increments. Fig. \ref{scale} shows the result for \textit{Kosarak} dataset with $min\_sup = 0.1\%$.
\textit{InUSP} and $WIncSpan'$ requires slightly more time at the initial point as they have to find and buffer the semi-frequent patterns for future use. 
After that, at any point of dataset increment, both of them take significantly less time to find the updated set of frequent sequences.
Our proposed technique outperforms the baseline approach in terms of scalability and although it takes slightly more time than $WIncSpan'$, the difference is negligible as \textit{InUSP} provides better completeness.\\
\textbf{(c) Varying Initial Size of Datasets: }
We considered different initial sizes for this analysis and introduced required number of increments (each sized 50-80\% of the initial size) to use the full dataset. Fig. \ref{comp_init} shows the result in \textit{Chainstore}
and \textit{Online Retail
dataset} with $min\_sup=0.05\%$  for both.
We have found that the
smaller the initial dataset, the more are the sequences to be found as new patterns after the increments.
The completeness of incremental approaches also depends on the distribution of items among the increments.
As a result, the completeness of $WIncSpan'$ is competitive only if the initial dataset contains sufficient sequences compared to the total size of all future increments.
However, the completeness of \textit{InUSP} is less affected by initial size as it also mines in the incremented portions. 

\section{Conclusions}
In this work, our proposed \textit{FUSP} algorithm can mine sequential patterns in uncertain databases with or without weight constraints. It uses multiple theoretically tightened upper bounds in pruning technique and hence, generates a smaller number of false-positive patterns compared to the state-of-the-art works. Furthermore, the use of a space-efficient data structure \textit{USeq-Trie} for pattern maintenance and an efficient method \textit{SupCalc} for support calculation, has made \textit{FUSP} superior to other works in terms of runtime.
In case of incremental mining, the concept of promising frequent sequences lifts the effectiveness of our \textit{InUSP} algorithm. 
The experimental analysis shows that our proposed techniques can be great tools for a lot of real-life applications such as medical records, sensor network, user behavior analysis, privacy-preserving data mining, that use uncertain sequential data.
We hope that the concept of \textit{USeq-Trie} structure and promising frequent sequences will help researchers to design efficient mining methods in related fields (e.g., uncertain data streams, spatio-temopral data, etc).\\
\newline
{\bf Acknowledgement.} This project is partially supported by NSERC (Canada) and University of Manitoba.

\bibliographystyle{splncs04}
\bibliography{ref.bib}
\end{document}